\documentstyle[12pt]{ioplppt}

\begin{document}
\jl{1}
\title{Directed percolation near a wall} 

\author{J W Essam\dag, A J Guttmann\ddag, 
I Jensen\ddag\ and D TanlaKishani\dag}

\address{\dag Department of Mathematics, Royal Holloway, 
University of London, Egham Hill, Egham, Surrey TW20 0EX, UK}

\address{\ddag Department of Mathematics,
University of Melbourne, Parkville, Victoria 3052, Australia}

\begin{abstract}
Series expansion methods are used to study directed bond percolation 
clusters on the square lattice whose lateral growth is restricted by 
a wall parallel to the growth direction. The percolation threshold 
$p_c$ is found to be the same as that for the bulk. However the 
values of the critical exponents for the percolation probability and 
mean cluster size are quite different from those for the bulk and 
are estimated by $\beta_1 = 0.7338 \pm 0.0001$ and 
$\gamma_1 = 1.8207 \pm 0.0004$ respectively. On the other hand the 
exponent $\Delta_1= \beta_1 + \gamma_1$ characterising the scale of 
the cluster size distribution is found to be unchanged by the 
presence of the wall.

The parallel connectedness length, which is the scale for the 
cluster length distribution, has an exponent which we estimate to be
$\nu_{1\parallel} = 1.7337 \pm 0.0004$ and is also unchanged. The 
exponent $\tau_1$ of the mean cluster length is related to $\beta_1$ 
and $\nu_{1\parallel}$ by the scaling relation 
$\nu_{1\parallel} = \beta_1 +  \tau_1$ and using the above estimates 
yields $\tau_1 = 1$ to within the accuracy of our results. 
We conjecture that this value of  $\tau_1$ is exact and further support
for the conjecture is provided by the direct series expansion estimate
$\tau_1= 1.0002 \pm 0.0003$.
\end{abstract}


\maketitle

Recently exact results have been obtained for directed compact 
clusters on the square lattice near a wall \cite{ET,EG,JCL}. 
Such clusters are similar to ordinary percolation clusters except 
that they cannot branch and have no holes. These simplifying features
allow several of the usual percolation functions to be derived 
analytically and the corresponding critical exponents have integer 
values. One of the  main conclusions from these results was that although 
the moments of the cluster size and length distributions have exponents 
which change on introducing a wall parallel to the growth direction 
the exponents for the size and length scales remain the same. The 
other was that growth parallel to the wall is rather special in 
that any bias away from the wall results in bulk exponents. Similarly 
any bias towards the wall leads to wet wall exponents \cite{ET}.

In this paper we find that the first of these conclusions extends to 
directed bond percolation. The exponents for directed percolation 
are not known exactly but numerical results show that, even in the 
absence of a wall, they are generally far from being integer
and there is some doubt as to whether they even have rational values
\cite{IGW}. 
An interesting possibility raised by our results is that the mean
cluster length in the presence of a wall parallel to the growth 
direction is an exceptional case and has the integer exponent 
$\tau_1 = 1$. Direct evidence for this value is provided by our 
analysis of the low density series expansion for the mean cluster 
length. Further support is provided by the scaling relation
\begin{equation}\label{e1}
	\beta_1 +  \tau_1 = \nu_{1\parallel}
\end{equation}
together with series expansion estimates of $\beta_1$ and 
$\nu_{1\parallel}$. Here the subscript $1$ on an exponent indicates 
its value in the presence of a wall. This relation is less well 
known than the one for the cluster size distribution, namely
\begin{equation}\label{e2}
	\beta_1 +  \gamma_1  = \Delta_1
\end{equation}
and is derived below. First we define the model and introduce some 
notation.

The directed square lattice may be described as having sites which 
are the points in the $t - x$ plane with integer co-ordinates 
such that $t \ge 0$ and $t + x$ is even. There are two bonds leading
from the general site $(t,x)$ which terminate at the sites 
$(t +1, x \pm 1)$. All bonds have probability $p$ of being open 
to the passage of fluid and the source is placed at $(0,0)$. This 
will be known as the bulk problem. A wall will be said to be 
present if the bonds leading to sites with $x < 0$ are always 
closed. The probability that fluid reaches column $t$ but no further 
will be denoted by $r_t(p)$ and in this event 
the origin will be said to belong to a cluster of length $t$.

The percolation probability, the probability that the origin 
belongs to a cluster of infinite length, is defined by
\begin{equation}\label{e3}
P(p) =1-\sum_{t=0}^\infty r_t(p) =  	 
       \sum_{t=0}^\infty (r_t(p_c) - r_t(p) )
\sim (p-p_c )^{\beta} \hbox{  for  } p \rightarrow p_c^+.
\end{equation}
If we suppose that the length distribution has the scaling form
\begin{equation}\label{e4}
r_t(p)  \sim  t^{ -a} f( t / \xi_\parallel(p) )
\end{equation}
then if
\begin{equation}\label{e5}
	\xi_\parallel(p) \sim	 | p_c - p |^{-\nu_\parallel}
\end{equation}
substitution in  (\ref{e3}) yields
\begin{equation}\label{e6}
		a = 1 + {{\beta}\over{ \nu_\parallel}}.
\end{equation}
The mean cluster length is defined by
\begin{equation}\label{e7}
T(p)     = 	 \sum_{t=0}^\infty  t r_t(p)
\end{equation}
and using (\ref{e4}) we find that
\begin{equation}\label{e8}
		T(p)   \sim  | p_c - p | ^{-\tau}
\end{equation}
where
\begin{equation}\label{e9}
		\tau =  \nu_\parallel - \beta.
\end{equation}
The same argument holds in the presence of the surface and leads 
to (\ref{e1}). There is a close correspondence between the above 
derivation and that of  (\ref{e2}) given in \cite{DS}. To obtain 
(\ref{e2}) it is only necessary to replace $r_t(p)$ by the 
cluster size distribution $p_s(p)$, $\xi_\parallel(p)$ by the scaling
size $\sigma(p)$, which diverges with critical exponent $\Delta$, 
and $T(p)$ by the mean cluster size $S(p)$ which diverges with 
exponent $\gamma$.

The mean size and the parallel and perpendicular scaling lengths are 
obtained from the pair connectedness function $C(t, x; p)$ which is 
the probablity that there is an open path from the origin to the
site $(t,x)$. The moments are defined by
\begin{equation}\label{e10}
\mu_{m,n}(p)       = \sum_{sites}  t^m x^n C(t, x; p)
\end{equation}
in terms of which $S(p) = \mu_{00}(p)$. Assuming a scaling form 
for $C(t, x; p)$ similar to (\ref{e4}),
where $x$ is scaled by $\xi_\perp(p)$, it follows that
\begin{equation}\label{e11}
\xi_\parallel(p)  \sim {{\mu_{m0}(p)} \over{ \mu_{m-1,0}(p)}}        
\hspace{0.5 in}\hbox{  and  } \hspace{0.5 in}  
\xi_\perp(p)    \sim   {{\mu_{0n}(p)}\over{ \mu_{0,n-1}(p)}}.
\end{equation}
The series expansions are obtained by a transfer matrix method similar 
to that used for the bulk lattice \cite{IJ} and the details of the
implementation in the presence of a wall will be given in 
a forthcoming paper \cite{IGW}. 
The state of column $t$ is a specification of which sites in that 
column are wet and which are dry and the probability that state
$i$ occurs is denoted by $\pi_i(t,p)$. The state in which all sites 
are dry is labelled  $i  = 0$. Essentially the state vector of a 
given column is completely determined by that of the previous 
column and only one state vector need be held in the computer 
at any stage. $C(t, x; p)$ is determined by summing $\pi_i(t,p)$ 
over all states for which the site with co-ordinate $x$ is wet and
\begin{equation}\label{e12}
		 r_t(p) = \pi_0(t+1,p) - \pi_0(t,p).
\end{equation}
Low density expansions in powers of $p$ are obtained by noting that 
$\pi(t,p) = {\cal O}(p^t)$ so that all of the above functions may be 
obtained to this order by computing the state vectors up to
column $t$. We were able to derive the series directly up to
a maximal column $t_m = 49$.
However, these series can be extended significantly
via an extrapolation method similar to that of \cite{BG}. 
As an example, consider the series for the average cluster length
$T(p)$. For each $t < t_m$ we calculate the polynomials 
$T_{t}(p) = \sum_{t'=0}^t t'r_{t'}(p)$ correct to ${\cal O}(p^{70})$.
As already noted these polynomials agree with the series for $T(p)$
to ${\cal O}(p^t)$. Next, we look at the sequences $d_{t,s}$ obtained
from the difference between successive polynomials
\begin{equation}\label{e13}
T_{t+1}(p)-T_t(p) = p^{t+1} \sum_{s \geq 0} d_{t,s} p^s.
\end{equation}
\noindent
The first of these correction terms $d_{t,0}$ is often a 
simple sequence which one can readily identify. In this case
we find the sequence
$$
-d_{t,0} = 1,2,3,6,10,20,35,70,126,252,462,924,\ldots
$$
\noindent
from which we conjecture
\begin{equation}\label{e14}
  d_{2t,0} = 2d_{2t-1,0}, \;\;\; d_{2t-1} = 4^t/B(t+1,-1/2),
\end{equation}
\noindent
where $B(x,y)$ is the Beta function. The formula for $d_{t,0}$ holds
for all the $t_m \! - \!1$ values that we calculated and we are 
very confident that it is correct for all values of $t$. As was the
case in \cite{BG} the higher-order correction terms $d_{t,s}$
can be expressed as rational functions of $d_{t,0}$,
\begin{equation}\label{e15}
  d_{t,s} = \sum_{k=1}^{[s/2]}
  \left( \begin{array}{c} t-s \\ k \end{array} \right)
  (a_{s,k} d_{t-s+1,0} + b_{s,k} d_{t-s+2,0})
  +\sum_{k=0}^{s} c_{s,k} d_{t-s+k+1,0}.
\end{equation}
\noindent
From this equation we were able to find formulas for
all correction terms up to $s=17$ and using $T_{49}(p)$ we
could extend the series for $T(p)$ to ${\cal O}(p^{67})$.
A similar procedure allowed us to extend the series for
$S(p)$ and the parallel moments $\mu_{1,0}(p)$ and
$\mu_{2,0}(p)$ to ${\cal O}(p^{67})$, while the series for
the first and second perpendicular moments, $\mu_{0,1}(p)$
and $\mu_{0,2}(p)$, were extended to ${\cal O}(p^{65})$.
The resulting series are listed in table \ref{series}. More details
of the extrapolation procedure including the formulae for
the various correction terms will appear in a later paper \cite{IGW}.

The only high density expansion we consider is that for the 
percolation probability which can be obtained from (\ref{e12}) 
and (\ref{e3}) by noting that $r_t(p) = O(q^k)$ where $q = 1-p$ 
and $k$ is the least integer $\ge \frac12 (t +2)$. Thus for a given 
value of $t$ the number of terms obtainable in the high density 
expansion is only about half as many as in the low density expansion. 
However, for computational purposes it is more efficient to
derive the series expansion for $P(q)$ directly via a transfer
matrix technique.
For the percolation probability we derived the series
directly to ${\cal O}(q^{24})$ and obtained another 8 terms
from the extrapolation procedure.
The resulting series is listed in table \ref{ppseries}.

It is found from unbiassed approximants that the estimates 
of $p_c$ agree with the bulk value \cite{IJ},  
$p_c = 0.6447002 \pm 0.0000005$ obtained from longer series and we 
therefore bias our exponent estimates with this value. This value of
$p_c$ was obtained from low density series and is a refinement of that 
obtained from analysis of the shorter series for $P(q)$ \cite{JG} 
which gave $p_c=0.6447006 \pm 0.0000010$. Data obtained
from $T(p)$, the parallel moments and $P(q)$ is shown in tables 
\ref{length}, \ref{moments} and \ref{percprob}. The exponent of
$\mu_{00}(p)$ was estimated from the series for $(S(p)-1)/p$ which is the 
mean size of the cluster connected to the site (1,1); 
this gave better convergence. 
We have also analysed the first and second perpendicular moment of the 
pair connectedness and series for $\xi_\parallel(p)$ and 
$\xi_\perp(p)$ obtained from (\ref{e11}) using the first and
second moments. In the analysis of
$P(q)$ we used standard DLog Pad\'e approximants while the remaining
series were analysed using first and second order inhomogeneous
differential approximant \cite{AJG}. 

In table \ref{length} the columns headed $L=0$ result from the 
standard DLog Pad\'e analysis and give $\tau_1 = 1$ to three decimal 
places although most of the entries are slightly above. This conclusion
is not altered by looking at inhomogeneous approximants (the first few of
which we have included in table \ref{length}) or second order 
approximants. Using the slightly smaller value  $p_c=0.6446980$ gave
the better converged result $\tau_1 = 1.00004 \pm 0.00004$.

We turn now to the indirect evidence for $\tau_1 = 1$ via the scaling
relation (\ref{e1}). The value $\nu_{1\parallel} = 1.7337\pm0.0004$ 
was obtained by analysing the series for $\mu_{2,0}(p)/\mu_{1,0}(p)$
and is consistent with the value obtained by subtracting the
value of the exponent of $\mu_{1,0}$ from that of $\mu_{2,0}$. It is 
clearly equal to the corresponding bulk exponent, as in the case of 
compact percolation, and we use the more accurate bulk estimate 
in deriving $\tau_1$ below. The corrections to scaling in the case of the 
percolation probability appear to be very close to analytic, 
and the standard Pad\'e estimate of $\beta_1$ 
(table \ref{estimates}) should be accurate. 
Combining the values of $\nu_{\parallel}$ and $\beta_1$ gives 
$\tau_1 = 1.0000 \pm 0.0002$ which agrees with the direct estimate. 

Other exponent values obtained from the analysis of various series
are collected together in table 
\ref{estimates} where previous estimates for the bulk problem and 
exact results for compact percolation are also given. As usual 
the error bars are a measure of the consistency of the higher order 
approximants and are not strict bounds. The estimate 
$\beta = 0.27643 \pm 0.00010$ of \cite{JG} has been adjusted slightly
upwards to allow for the change in $p_c$.
In estimating the exponents
we rely both on the analysis of the series yielding a particular
exponent and estimates obtained using scaling relations. In some cases
we also use the more accurate bulk exponent estimates. A case in point
is the exponent $\gamma_1$. From the Dlog Pad\'e approximants in
table \ref{moments} one would say that the direct estimate from
the series for $(S(p)-1)/p$ favours a value of 
$\gamma_1 \simeq 1.8211$ with
a rather large spread among the approximants. However,
the better converged estimates of 
$\gamma_1 +2\nu_{1\parallel} \simeq 5.2881$
together with the bulk estimate of $\nu_\parallel$ leads
to $\gamma_1 \simeq 1.8205$. In this case second order differential
approximants to $S(p)$ are better converged and favour 
$\gamma_1 \simeq 1.8207$. Taking all the evidence into account
including our belief that $\Delta_1$ takes on the bulk value we
arrived at the estimate for $\gamma_1$ quoted in table \ref{estimates}.
The estimate of $\tau$ is derived from the scaling relation (\ref{e9}).
Analysis of the bulk expansions \cite{IJ,EDAB,JG} showed 
that corrections to scaling were close to analytic, as they are here. 

The values of $\nu_{1\perp}$ and $\Delta_1$ (obtained from the scaling 
relation (\ref{e2})), as well as $\nu_{1\parallel}$, are clearly the
same as those for the bulk. The scaling size and both scaling lengths
are therefore unchanged by the introduction of the wall. We also note that
the hyperscaling relation, with
$D$ the dimension of space perpendicular to the preferred 
direction $t$ (=1 for the square lattice),
 \begin{equation}\label{e16}
\nu_\parallel +  D\nu_\perp = \beta +  \Delta,
\end{equation}  
which is satisfied by the bulk exponents apparently
fails on the introduction of a wall. 

We now consider the possibility of rational exponents. As previously noted
\cite{JG}, there is no simple rational
fraction whose decimal expansion agrees with the estimate of $\beta$.
The same is true for other exponent estimates in table \ref{estimates}.
In particular we note that our estimates of the bulk exponents
$\nu_{||}$ and 
$\nu_{\perp}$ differ by $0.03\%$ from the rational fractions
$\nu_{||} = 26/15 = 1.733\; 333 \ldots$,
and $\nu_{\perp} = 79/72 = 1.097\; 222 \ldots$ suggested by
Essam \etal \cite{EGD}. We believe this to be
a significant difference given the high precision of our results.
However, the suggested rational fraction
$\gamma = 41/18 = 2.277\; 777 \ldots$ and the value of
$\Delta = 613/240 = 2.554\; 1666 \ldots$, which follows from the above
rational values by scaling, are generally still within our estimated
the error bounds. The fraction for $\Delta$ is
not very appealing though and assuming that both
exponents have these values then scaling implies the even less convincing
result $\beta = 199/720 = 0.276\; 388 \dots$ which is however just 
consistent with our estimated value. 

If we assume that $\tau_1 = 1$ is exact and that the values of $\Delta$, 
$\nu_\parallel$ and $\nu_\perp$ are the same with and without a wall then
all of the other surface exponents are determined by scaling together 
with the values of any three 
bulk exponents. The surface exponents calculated in 
this way are presented in table \ref{scaling} for comparison with the 
estimated values of table \ref{estimates} as a measure of the overall 
consistency of our results. The bulk exponents used were $\gamma = 41/18$
and the bulk estimates of $\nu_{||}$ and $\nu_{\perp}$.
Excellent agreement is observed.

Our findings may be summarised as follows. Firstly we have found that the
scaling size and both scaling length exponents are unchanged by the
introduction of a wall parallel to the preferred direction. 
Also we have examined the widely held view that two dimensional systems 
should have rational exponents. The high precision data 
presented here is consistent with the results $\tau_1 = 1$ and
$\gamma = 41/18$. However there are no such simple fractions which
are in agreement with our estimates of $\nu_{||}$ and $\nu_{\perp}$.
Given that directed percolation is not conformally invariant,
and that the expectation of
exponent rationality is a consequence of conformal invariance, this
is perhaps not surprising.
The precise numerical work reported and quoted 
in this paper therefore supports the conclusion that the critical
exponents for non translationally invariant
models should not in general be expected to be simple rational numbers. 
 The cluster length exponent $\tau_1$ and the
exponent $\gamma$ appear to be exceptional cases.

\Tables

\hoffset=-1cm
\begin{table}
\caption{Low density expansions in powers of $p$, row $n$ 
is the coefficient of $p^n$.\label{series}}
\renewcommand{\baselinestretch}{0.75}
\tiny
\begin{tabular}{rrrrrrr} 
\br
$n$ & $T(p)$ & $S(p)$ & $\mu_{1,0}(p)$ & $\mu_{2,0}(p)$ 
& $\mu_{0,1}(p)$ & $\mu_{0,2}(p)$\\
\mr
0 & 0 & 1 & 0 & 0 & 0 & 0 \\
1 & 1 & 1 & 1 & 1 & 1 & 1 \\
2 & 2 & 2 & 4 & 8 & 2 & 4 \\
3 & 2 & 3 & 9 & 27 & 5 & 11 \\
4 & 5 & 6 & 24 & 96 & 10 & 28 \\
5 & 5 & 9 & 47 & 241 & 21 & 65 \\
6 & 11 & 17 & 108 & 672 & 40 & 144 \\
7 & 13 & 26 & 201 & 1499 & 77 & 303 \\
8 & 28 & 47 & 424 & 3676 & 142 & 624 \\
9 & 25 & 72 & 762 & 7644 & 262 & 1240 \\
10 & 75 & 129 & 1538 & 17398 & 470 & 2438 \\
11 & 56 & 194 & 2675 & 34369 & 843 & 4661 \\
12 & 188 & 348 & 5258 & 74512 & 1486 & 8872 \\
13 & 112 & 516 & 8915 & 141615 & 2609 & 16487 \\
14 & 458 & 929 & 17233 & 296939 & 4529 & 30635 \\
15 & 319 & 1351 & 28518 & 546394 & 7846 & 55734 \\
16 & 1157 & 2456 & 54636 & 1119562 & 13448 & 101618 \\
17 & 312 & 3506 & 88459 & 2004015 & 23027 & 181751 \\
18 & 3389 & 6471 & 169004 & 4043156 & 39096 & 326608 \\
19 & 562 & 8929 & 266670 & 7047626 & 66320 & 575790 \\
20 & 9193 & 17029 & 512651 & 14102481 & 111795 & 1022909 \\
21 & -2419 & 22579 & 786932 & 23956166 & 187946 & 1781314 \\
22 & 24689 & 44707 & 1530464 & 47809422 & 314844 & 3135130 \\
23 & -6090 & 55969 & 2270857 & 79011279 & 526367 & 5402999 \\
24 & 83997 & 117836 & 4516598 & 158359672 & 876362 & 9435440 \\
25 & -80845 & 137313 & 6439085 & 254037643 & 1455579 & 16106911 \\
26 & 219791 & 311654 & 13207919 & 514524887 & 2415059 &
 27970523 \\
27 & -95543 & 324989 & 17852082 & 796972392 & 3989542 &
 47305236 \\
28 & 653560 & 833496 & 38438680 & 1646320650 & 6597538 &
 81807186 \\
29 & -1015961 & 756309 & 48640815 & 2447308375 & 10834513
 & 137158135 \\
30 & 2302634 & 2242031 & 111440275 & 5201705453 & 17869253
 & 236510661 \\
31 & -2111933 & 1623709 & 128688532 & 7341847456 & 29239356
 & 393079288 \\
32 & 6978051 & 6176873 & 324010503 & 16294292667 & 48152477
 & 677071243 \\
33 & -12164131 & 3240757 & 331752781 & 21552447211 & 
  78162313 & 1114451899 \\
34 & 21361373 & 17192674 & 944134956 & 50707490638 & 
  128852132 & 1921593186 \\
35 & -27110387 & 4663165 & 810982473 & 61539314001 & 
  208370375 & 3130415149 \\
36 & 93655507 & 49481888 & 2781591612 & 157488162524 & 
  343409668 & 5411807564 \\
37 & -182370254 & 1180046 & 1866117373 & 170712205993 & 
  549693819 & 8710776761 \\
38 & 229034090 & 144593684 & 8270004945 & 489038638889 & 
  911531157 & 15152834441 \\
39 & -269557768 & -40561669 & 3647454015 & 452466460859 & 
  1447853041 & 24030119951 \\
40 & 1056409556 & 439929287 & 25083883563 & 1526926232817 &
  2413312231 & 42187579545 \\
41 & -2269021879 & -230303695 & 5007776568 & 1132548161360 &
  3773060280 & 65731749816 \\
42 & 2677408443 & 1351358555 & 77130163183 & 4798086858971 &
  6361278369 & 117017657827 \\
43 & -3544761784 & -1116634980 & -6211741855 & 2514662834523 
  & 9833452727 & 178182324707 \\
44 & 13082866127 & 4353263697 & 244028578766 & 15284660803552
  & 16833476130 & 323726387136 \\
45 & -26806541805 & -4398416071 & -83631438989 & 4380744364749
  & 25157427559 & 478236033969 \\
46 & 26061243131 & 14001291871 & 783204867296 & 49292061993412
  & 44287084338 & 894531996536 \\
47 & -40361968343 & -17738446374 & -494314396278 & 989931047506
  & 64933486366 & 1270849732090 \\
48 & 190465471378 & 47119949250 & 2594611285466 & 
  162241456668132 & 117606789796 & 2471975021852 \\
49 & -381128060099 & -64270709097 & -2232294549879 & 
  -39805018765919 & 161582598415 & 3328670679553 \\
50 & 225643036457 & 157128098347 & 8690778026386 & 
  542342994602556 & 311756741490 & 6859481787132 \\
51 & -287003337097 & -246380178827 & -9661864892692 & 
  -284699866038824 & 408491249744 & 8579303387168 \\
52 & 2566759769655 & 545460020544 & 29995760431218 & 
  1856106540303732 & 841943528892 & 19102460884304 \\
53 & -5285267101147 & -862856345434 & -38056677957915 & 
  -1431588334552263 & 968313512109 & 21611403485081 \\
54 & 2271123259017 & 1858869421298 & 103906790631563 & 
  6441871877547593 & 2256308657115 & 53709860916525 \\
55 & -3165468030218 & -3252844644627 & -151969740070893 & 
  -6562243329132823 & 2354715740977 & 52606208892861 \\
56 & 35212809299763 & 6592890548347 & 369827081677281 & 
  22869643990253339 & 6364532607737 & 152781299898183 \\
57 & -66427001953763 & -11229139704329 & -570503946433867 & 
  -27580998453503811 & 4823911367581 & 121017115594937 \\
58 & 11057548952493 & 22767401371634 & 1310843427572251 & 
  81922344320438959 & 17432800454267 & 441260107224351 \\
59 & -31697059334297 & -42147789558521 & -2209141231427900 & 
  -114301635466580028 & 10767177749158 & 253668652604268 \\
60 & 531845697600814 & 81707816765666 & 4757125831653685 & 
  299099704878008319 & 52298853703005 & 1298380307866003 \\
61 & -939850501378691 & -144224611556818 & -8109804036235413 & 
  -452153132335049221 & 10274067757479 & 411221700812127 \\
62 & -218089303232488 & 284988594853047 & 17109904775959109 & 
  1095748251643358129 & 149825804840191 & 3920538018919121 \\
63 & 146310515780374 & -544069973568349 & -31055984288473750 & 
  -1802157080659641406 & 3194083769764 & 164257826455782 \\
64 & 8010088501049393 & 1029622326675184 & 62805743084099736 & 
  4074933118400663972 & 488096955080292 & 12077039640386216 \\
65 & -13777249481066198 & -1844661752754855 & -112541611208180874
   & -6931655775629313164 & -219315581678014 & -3036358866297604 \\
66 & -7335657891417937 & 3612493459852700 & 227780508663102551 &
   15135810090250397585 &  &  \\
67 & 5810530478862470 & -7025211744800954 & -429949623442589455 &
    -27153914600589832779 &  &  \\
\br
\end{tabular}
\end{table}

\newpage
\hoffset=0cm
\renewcommand{\baselinestretch}{1.0}

\Table{High density expansion for the percolation probability
$P(q) = \sum a_nq^n$.}
\br
\scriptsize
$n$ & $a_n$ &$n$ & $a_n$ \\
\mr
0 & 1  & 17 & -123721 \\
1 & -1  & 18 & -287828 \\
2 & -2  & 19 & -790641 \\
3 & -3  & 20 & -1875547 \\
4 & -4  & 21 & -5302725 \\
5 & -7  & 22 & -12258340 \\
6 & -11  & 23 & -35837868 \\
7 & -24  & 24 & -83642760 \\
8 & -44  & 25 & -242399471 \\
9 & -108  & 26 & -569416045 \\
10 & -221  & 27 & -1704989414 \\
11 & -563  & 28 & -3898028574 \\
12 & -1234  & 29 & -11682423741 \\
13 & -3240  & 30 & -28476236374 \\
14 & -7221  & 31 & -80448369426 \\
15 & -19835  & 32 & -194172723271 \\
16 & -44419  & & \\
\br
\label{ppseries}
\endTable

\Table{\label{length}
Differential approximant analysis of the mean length series.
The table shows biased first order inhomogeneous approximant 
estimates of $\tau_1$. $L$ is the degree of the inhomogeneous
polynomial. For $L=0$ the entries are from biased Dlog Pad\'e
approximants.}
\br
 & \multicolumn{3}{c}{$L = 0 $} &
 \multicolumn{3}{c}{$L = 1 $} \\
 & \crule{3} & \crule{3} \\
 N &  [N--1,N] & \centre{1}{[N,N]} & [N+1,N] 
  &  [N--1,N] & \centre{1}{[N,N]}  & [N+1,N] \\ \mr
 22 & 1.00010  & 1.00010  & 1.00010  
  & 1.00010 & 1.00014 & 1.00010 \\
 23 & 1.00010  & 1.00010  & 1.00009    
  & 1.00007 & 0.99923 & 1.00012 \\
 24 & 1.00010  & 1.00015  & 1.00019    
  & 1.00012 & 1.00006 & 1.00015 \\
 25 & 1.00019  & 1.00008  & 1.00021    
  & 1.00015 & 1.00016 & 1.00019 \\
 26 & 1.00021  & 1.00020  & 1.00021    
  & 1.00019 & 1.00020 & 1.00006 \\
 27 & 1.00021  & 1.00012  & 1.00027    
  & 1.00010 & 1.00025 & 1.00033 \\
 28 & 1.00028  & 1.00024  & 1.00034    
  & 1.00036 & 1.00023 & 1.00072 \\
 29 & 1.00001  & 1.00020  & 1.00024    
  & 1.00010 & 1.00020 & 1.00021 \\
 30 & 1.00028  & 1.00022  & 1.00022    
  & 1.00021 & 1.00022 & 1.00026 \\
 31 & 1.00022  & 1.00029  & 1.00023    
  & 1.00026 & 1.00025 & 1.00023 \\
 32 & 1.00023  & 1.00022  & 1.00022    
  & 1.00023 & 1.00022 & 1.00022 \\
 33 & 1.00022  & 1.00022  &  
  & 1.00022 &  &  \\  
  \mr
 & \multicolumn{3}{c}{$L = 2 $}   &
 \multicolumn{3}{c}{$L = 3 $} \\ 
 & \crule{3} & \crule{3} \\
 N & [N--1,N] & \centre{1}{[N,N]} & [N+1,N] 
  & [N--1,N] & \centre{1}{[N,N]} & [N+1,N] \\ \mr
 22 & 1.00010 & 1.00001 & 0.99954 
  & 1.00010 & 0.99996 & 1.00014 \\
 23 & 0.99962 & 1.00004 & 1.00009 
  & 1.00014 & 1.00008 & 1.00011 \\
 24 & 1.00009 & 0.99996 & 1.00014 
  & 1.00012 & 1.00017 & 1.00015 \\
 25 & 1.00014 & 1.00017 & 1.00144 
  & 1.00015 & 1.00019 & 1.00034 \\
 26 & 0.99986 & 1.00028 & 1.00030 
  & 1.00038 & 1.00030 & 1.00029 \\
 27 & 1.00030 & 1.00028 & 0.99995 
  & 1.00030 & 1.00029 & 1.00018 \\
 28 & 1.00014 & 1.00020 & 1.00020 
  & 1.00019 & 1.00020 & 1.00022 \\
 29 & 1.00020 & 1.00020 & 1.00025 
  & 1.00023 & 1.00022 & 1.00022 \\
 30 & 1.00030 & 1.00022 & 1.00023 
  & 1.00022 & 1.00023 & 1.00023 \\
 31 & 1.00023 & 1.00023 & 1.00022 
  & 1.00023 & 1.00022 & 1.00023 \\
 32 & 1.00023 & 1.00023 &  
  & 1.00023 &  &  \\  
\br
\endTable

\fulltable{\label{moments}
DLog Pad\'e analysis of the moments of the pair connectedness.
The table shows biased approximant estimates of the critical exponents 
of the moments $\mu_{00}(p)$, $\mu_{10}(p)$ and $\mu_{20}(p)$.}
\br
 & \multicolumn{3}{c}{$\gamma_1$} &
 \multicolumn{3}{c}{$\gamma_1+\nu_{1\parallel}$} &
 \multicolumn{3}{c}{$\gamma_1+2\nu_{1\parallel}$} \\ 
  & \crule{3} & \crule{3} & \crule{3} \\
 N & [N--1,N] & \centre{1}{[N,N]} & [N+1,N] 
   & [N--1,N] & \centre{1}{[N,N]} & [N+1,N] 
  & [N--1,N] & \centre{1}{[N,N]} & [N+1,N] \\ \mr
 22 & 1.82381  & 1.82760  & 1.81953  
  & 3.55492  & 3.55555  & 3.55458   
  & 5.28807  & 5.28807  & 5.28808  \\  
 23 & 1.82010  & 1.82593  & 1.82355    
  & 3.55466  & 3.55478  & 3.55472    
  & 5.28809  & 5.28769  & 5.28807  \\  
 24 & 1.82364  & 1.82094  & 1.71766    
  & 3.55473  & 3.55479  & 3.55462    
  & 5.28808  & 5.28804  & 5.28809  \\  
 25 & 1.77437  & 1.81558  & 1.82399    
  & 3.55466  & 3.55456  & 3.55457    
  & 5.28809  & 5.28809  & 5.28809  \\  
 26 & 1.82511  & 1.82793  & 1.82424    
  & 3.55457  & 3.55456  & 3.55459    
  & 5.28805  & 5.28809  & 5.28809  \\  
 27 & 1.82524  & 1.82063  & 1.82122    
  & 3.55460  & 3.55459  & 3.55459    
  & 5.28809  & 5.28808  & 5.28806  \\  
 28 & 1.82124  & 1.82097  & 1.82078    
  & 3.55460  & 3.55459  & 3.55459    
  & 5.28807  & 5.28809  & 5.28516  \\  
 29 & 1.82079  & 1.82090  & 1.82396    
  & 3.55460  & 3.55453  & 3.55458    
  & 5.28804  & 5.28802  & 5.28819  \\  
 30 & 1.81878  & 1.82098  & 1.82107    
  & 3.55459  & 3.55458  & 3.55459    
  & 5.28804  & 5.28806  & 5.28799  \\  
 31 & 1.82108  & 1.82108  & 1.82107    
  & 3.55473  & 3.55454  & 3.55463    
  & 5.28805  & 5.28761  & 5.28799  \\  
 32 & 1.82108  & 1.82104  & 1.82106    
  & 3.55425  & 3.55471  & 3.55460    
  & 5.28805  & 5.28806  & 5.28808  \\  
 33 & 1.82106  & 1.82104  &   
  & 3.55589  & 3.55471  &    
  & 5.28803  & 5.28806  &   \\  
 \br 
\endfulltable

\Table{\label{percprob}
DLog Pad\'e analysis of the percolation probability series.
The table shows biased approximant estimates of $\beta_1$.}
\br
 N  & [N--1,N] & \centre{1}{[N,N]} & [N+1,N] \\ \mr
  8 & 0.73406  & 0.73406  & 0.73408  \\  
  9 & 0.73409  & 0.73409  & 0.73408  \\  
 10 & 0.73409  & 0.73403  & 0.73369  \\  
 11 & 0.73389  & 0.73388  & 0.73385  \\  
 12 & 0.73389  & 0.73381  & 0.73382  \\  
 13 & 0.73382  & 0.73381  & 0.73382  \\  
 14 & 0.73383  & 0.73381  & 0.73382  \\  
 15 & 0.73382  & 0.73382  & 0.73379  \\  
 16 & 0.73380  & 0.73382  &  \\   
\br 
\endTable

\Table{\label{estimates}
Exponent values for compact and bond percolation. The bulk values 
for bond percolation are from \protect{\cite{IJ}} except for 
$\beta$ which is from \protect{\cite{JG}}, adjusted for a small change
in $p_c$. The compact percolation 
results are from \protect{\cite{EG}} and references therein.
Values in brackets are obtained from scaling formulae. The ``with wall''
value of $\gamma$ is from second order differential approximants.}
\br
&\multicolumn{2}{c}{bond percolation}&
\multicolumn{2}{c}{compact percolation}\\
exponent&with wall&bulk&\m with wall&bulk\\
\mr
$\tau$ & 1.0002$\pm$0.0003 & (1.4573$\pm$0.0002) &\m 0 & 1 \\
$\beta$	& 0.7338$\pm$0.0001 & 0.27647$\pm$0.00010 &\m 2 &	1 \\
$\gamma$ &1.8207$\pm$0.0004& 2.2777$\pm$0.0001	&\m 1 & 2 \\
$\gamma+\nu_\parallel$ & 3.5546$\pm$0.0002&4.0113$\pm$0.0003 &\m (3) & (4) \\
$\gamma+2\nu_\parallel$ & 5.2881$\pm$0.0002 & 5.7453$\pm$0.0004&\m (5) & (6) \\
$\nu_\parallel$ & 1.7337$\pm$0.0004 & 1.7338$\pm$0.0001 &\m (2) & 2 \\
$\gamma+2\nu_\perp$ & 4.014$\pm$0.002&  4.4714$\pm$0.0004 &  & (3) \\
$\nu_\perp$ & 1.0968$\pm$0.0003 & 1.0969$\pm$0.0001 &  & 1 \\
$\Delta$ & (2.5545$\pm$0.0005) & (2.5542$\pm$0.0002) &\m 3 & 3 \\
\br 
\endTable

\Table{\label{scaling}
Scaling values of the exponents for bond percolation calculated using 
$\tau_1 =1$, $\gamma = {{41} \over {18}}$ and the bulk estimates of 
$\nu_\parallel$ and $\nu_\perp$. }
\br
exponent&with wall&bulk\\
\mr
$\tau$ & 1         &1.4573\\
$\beta$	& 0.7338    &0.27646 \\
$\gamma$ &1.8204   &2.2778	\\
$\gamma+\nu_\parallel$ & 3.5542   &4.0116\\
$\gamma+2\nu_\parallel$ &5.2880   &5.7454\\
$\nu_\parallel$ &1.7338    &1.7338\\
$\gamma+2\nu_\perp$ & 4.0142   &4.4716\\
$\nu_\perp$ &1.0969    &1.0969\\
$\Delta$ & 2.5542   & 2.5542\\
\br 
\endTable

\end{document}